\input harvmac

\Title{HUTP-98/A075}{Evading the GZK Cosmic--Ray Cutoff}
\centerline{Sidney Coleman and Sheldon L. Glashow\footnote{$^\dagger$}
{{\tt coleman@physics.harvard.edu, \ glashow@physics.harvard.edu}}}
\bigskip\centerline{Lyman Laboratory of Physics}
\centerline{Harvard University}\centerline{Cambridge, MA 02138}

\vskip .3in Explanations of the origin of ultra-high energy cosmic rays are
severely constrained by the Greisen-Zatsepin-Kuz'min effect, which limits
their propagation over cosmological distances. We argue that possible
departures from strict Lorentz invariance, too small to have been detected
otherwise, can affect elementary-particle kinematics  so as 
to suppress or forbid inelastic collisions of  cosmic-ray nucleons with
background photons. Thereby can the GZK cutoff be relaxed or removed.

\Date{08/98} 

Soon after the discovery of the cosmic background radiation (CBR),
Greisen~\ref\rgr{K. Greisen, Phys. Rev. Lett. {\bf 16} (1966) 748.}, and
independently,  Zatsepin amd Kuz'min~\ref\rkz{G.T. Zatsepin and V.A.
Kuz'min, JETP Lett. {\bf 41} (1966) 78.},  realized that it limits the
propagation of ultra-high energy (UHE) nucleons. Cosmic-ray  nucleons with
sufficient energy can suffer inelastic impacts with photons of the CBR.
This  results in what is known as the GZK cutoff, saying that nucleons
with energies over $5\times10^{19}\;$eV cannot reach us from further
than a few dozen Mpc. However, the cosmic-ray
energy spectrum has been observed to
extend well beyond this energy~\ref\takeda{M. Takeda,
Phys. Rev. Lett. {\bf 81} (1998) 1163.}.

The mechanism producing UHE cosmic rays is unknown. Various exotic origins
have been proposed, among them: topological defects, active galactic
nuclei,  and gamma-ray bursts~ \ref\rother{See \takeda\ and V. Berezinsky,
{\tt hep-th/9802351} for further references in these connections.}.  These
schemes are severely constrained, if not ruled out,  by the GZK cutoff.
Another explanation, that the most energetic cosmic rays are decay
products  of  hypothetical super-heavy relic particles~\ref\bere{V.
Berezinsky, M. Kachelrie\ss, and V. Vilenkin, Phys. Rev. Lett. {\bf 79}
(1997) 4302.} is consistent with  observations of six events by AGASA with
energies $>10^{20}$~eV that are widely separated in position and not
coincident with any known astrophysical source~\takeda.

We have little to say about the origin of UHE cosmic rays
{\it per se.} Rather, we point out that  there may not be a GZK
cutoff after all. 
Tiny departures from Lorentz invariance, too small to
have been detected otherwise, have effects that
increase rapidly with energy and can 
kinematically prevent cosmic-ray nucleons from
undergoing inelastic collisions with CBR photons.
If the cutoff is thereby  undone,  a
deeply cosmological origin of UHE cosmic rays would  become tenable.

We have no reason to anticipate a failure of special relativity nor 
are we aware of
any reasonable theory predicting one. Nonetheless, strict
Lorentz invariance should not be accepted 
on faith but rather as a plausible hypothesis subject
to experimental test. It follows that  we cannot accept the GZK cutoff as an
indisputable fact. Further observations of UHE cosmic rays
could confirm a predicted `bump' just below the cutoff~\ref\schr{C.T. Hill
and D.N. Schramm, Phys. Rev. {\bf D31} (1985) 564.}\ resulting
from products of inelastic collisions of primary protons with
CBR photons. This would prove the GZK cutoff to be at least partially
effective. Or, they could indicate sources at cosmological distances that
would belie the cutoff. 

A perturbative framework has been developed~\ref\rcg{S. Coleman and S. L.
Glashow,  Phys. Lett. {\bf B405} (1997) 249,
and work in progress; S. L. Glashow, Nucl. Phys. (Proc.
suppl.) {\bf B70} (1999) 180.  See also: S. L. Glashow
in Proc. Intl. School of Subnuclear Physics,
Erice '97, and  in Proc. Tropical
Conference '98 (Univ. Puerto-Rico) both to be publ.}~\ref\rk{See also: D.
Colladay and V.A. Kosteleck\'y, Phys. Rev. {\bf D57} (1997) 3932.}
from which to explore the
observable consequences of a tiny failure of special relativity.
Renormalizable and gauge-invariant perturbations to the standard-model
Lagrangian that are rotationally invariant in a preferred frame, {\it but
not Lorentz invariant,}  lead to species-specific   maximum attainable
velocities (MAV)
for different particles.\footnote{$^*$}{In general,  the
MAV can be both flavor-changing and parity-violating, but these
complications may be ignored for the present considerations.}
We designate by $c_a$ the MAV of a given particle species $a$.
The MAV's of leptons and photons appear explicitly in the
perturbed Lagrangian, while those of hadrons are implicit functions of 
parameters relating to quarks and gluons.
Our arguments are phrased in terms of the
the following measures of Lorentz violation:
$$\delta_{ab}= -\delta_{ba}\equiv c_a^2-c_b^2\,.$$
The velocity of the 
lab frame (Earth) relative to the preferred frame (plausibly
that in which the CBR is isotropic) is assumed non-relativistic
so that it may be ignored in our kinematic considerations.

To see how the GZK cutoff is affected by possible
violations of Lorentz invariance, we begin by considering the formation of
the first pion-nucleon resonance: 
\eqn\eform{ p+ \gamma\ {\rm(CBR)}\longrightarrow \Delta(1232)\,,} 
by a  proton of energy 
$E$ colliding with a CBR photon of energy $\omega$.  The target photon
energies have
a thermal distribution with temperature $T=2.73\;$K, corresponding to 
$\omega_0=2.35\times 10^{-4}\;{\rm eV}^2$. Energy conservation provides the
condition under which reaction \eform\ can proceed:  
\eqn\eforma{4\omega\ge\delta_{\Delta p}E+ {M_\Delta^2-M_p^2\over E}\,.} 
If Lorentz symmetry is
unbroken, $\delta_{\Delta p}=0$ and Eq.~\eforma\ yields the 
conventional threshold for a head-on impact: $E_f=
(M_\Delta^2-M_p^2)/4\omega$. Otherwise 
inequality \eforma\ is a quadratic form in $E$, satisfiable if and only if
$\delta_{\Delta p}\le \omega/E_f$. As $\delta_{\Delta p}$ increases from
zero,  the threshold grows toward $2E_f$ as
$\delta_{\Delta p}$ approaches its  critical value: 
\eqn\edelcrit{ \delta_{\Delta
p}= \hat{\delta}(\omega)\equiv {4\omega^2\over M_\Delta^2-M_p^2} \simeq 3.5\times
10^{-25}\;[\omega/\omega_0]^2\;.} 
{\it For $\delta_{\Delta p}>\hat \delta$, reaction \eform\ is kinematically
forbidden for all $E$.} Recalling that the photon spectrum is thermal, we
see that for $\delta_{\Delta p}$ comparable to $\hat{\delta}(\omega_0)$, the GZK
cutoff due to resonant $\Delta(1232)$ formation would be relaxed.
Should it much exceed this value, formation would be prevented 
off virtually all CBR photons

Reaction \eform\ is the dominant process leading to the GZK
cutoff, as  argued by Greisen,
Zatsepin and Kuz'min. However, if $\Delta(1232)$ formation is not
possible, a
 weakened version of the GZK cutoff
may result from non-resonant photo-production of one or more pions:
\eqn\eprod{p+ \gamma\ {\rm (CBR)}\longrightarrow p + N\,\pi\,.}
Ordinarily (for $\delta_{\pi p}=0$), 
the threshold for single pion production is
$E_p=M_\pi(2M_p+M_\pi)/4\omega$.  If Lorentz invariance is
violated and  $\delta_{\pi p}$ is imagined to increase,
the  threshold grows. For a fixed photon energy $\omega$, we
have shown~\rcg\ that the threshold diverges
as
$\delta_{\pi p}$ approaches the critical value:
\eqn\eproda{ \delta_{\pi p}=
{\tilde{\delta}(\omega)\over N^2}\equiv {4\omega^2\over N^2\, M_\pi^2}\simeq
1.1\times 10^{-23}\ [\omega/\omega_0]^2\,. }
For larger values of $\delta_{\pi p}$,
reaction \eprod\ is 
kinematically forbidden {\it at all proton energies.} 
For the actual case of a thermal distribution of photon energies, values of
$\delta_{\pi p}$ comparable to or greater than    $\tilde \delta
(\omega_0)$ would suppress 
photo-pion production, or even eliminate it entirely 
so that no vestige of the GZK
cutoff survives. 

We have shown that
tiny values of the 
{\it a priori\/} unknown Lorentz-violating
parameters $\delta_{\Delta p}$ and $\delta_{\pi p}$ can suppress or forbid 
the processes underlying the GZK cutoff. 
Note that much larger (and experimentally intolerable) violations of
Lorentz invariance are needed to affect the interactions of UHE
cosmic rays with nuclei in the atmosphere. For example, to forbid
the process $p+p\rightarrow p+p+N\pi$ would require $\delta_{\pi
p}> (M_p/NM_\pi)^2$.

Ther are no direct experimental constraints on the parameters
$\delta_{\Delta p}$ and $\delta_{\pi p}$. However, it may be instructive
to compare the MAV differences necessary to affect the
GZK cutoff with  current experimental constraints on various analogous
Lorentz-violating parameters. Here is  a list of the strongest constraints
we can identify: 
\eqna\econst 
$$\eqalignno{ |\delta_{m \gamma}|< &\ 
6\times 10^{-22} & \econst a\cr \delta_{p \gamma}  < &\  3\times 10^{-23}  
& \econst b \cr |\delta_{\nu\nu'}| < &\  2\times 10^{-21} & \econst c\cr
\delta_{\gamma e}  < &\  5\times 10^{-16}   & \econst d\cr |\delta_{\nu
\gamma}|< &\ 2\times 10^{-16}   & \econst e \cr }$$ 
None of these constraints reach the level of sensitivity to
Lorentz violation  needed to affect the GZK cutoff. Constraint \econst{a}\
results from a Hughes-Drever type experiment with $\delta_{m \gamma}\equiv
c_m^2-c_\gamma^2$ where $c_m$, the MAV of  material matter,  was taken to
be the same for all massive particles~\ref\rl{S.K. Lamoreaux {\it et al.,}
Phys. Rev. Lett. {\bf 57} (1986) 3125.}. Constraints \econst{b,c,d} have
been discussed elsewhere~\rcg. The one-sided constraint \econst{b}\ results
from alleged detections of primary protons with energies up to $2\times
10^{20}\;{\rm eV}$, which otherwise would have lost their energy via vacuum
Cerenkov radiation. The weaker one-sided constraint \econst{d}\  results
from  alleged detections of primary photons with energies up to 50~TeV,
which otherwise would have decayed into electron-positron pairs. Constraint
\econst{c}\ reflects the failure to detect velocity oscillations of
neutrinos. Finally, constraint \econst{e}\ {\it could be obtained\/} from
observations of neutrinos from gamma-ray bursts at cosmological
distances~\ref\jb{E. Waxman and J. Bahcall, Phys. Rev. Lett. {\bf 78}
(1997) 2292.}. Absent these data, the current and much weaker  limit $
|\delta_{\nu \gamma}| < 10^{-8}$  arises from the detection of neutrinos
from supernova~1987a~\ref\rsto{L. Stodolsky, Phys. Lett. {\bf B201} (1988)
353.}.

Existing bounds on departures from special relativity are insufficiently
precise to disfavor those required to mitigate the GZK cutoff. Fortunately,
several bounds can be strengthened so as to approach the values indicated
in Eqs.~\edelcrit\ and \eproda.  Laboratory tests of Lorentz invariance far
more precise than any done before are now feasible~\ref\rfort{E.N. Fortson,
private communication.}. Dedicated searches for velocity oscillations of
solar neutrinos, or of accelerator-produced $\sim\,$TeV neutrinos at
baselines of $\sim\! 1000$~km,  can reveal neutrino velocity differences as
small as $10^{-25}$. At present, lacking detailed observations of the
highest energy cosmic rays and more precise tests of special relativity, we
must regard   as intriguingly open questions  both the existence of the GZK
cutoff and a cosmologically remote  origin of UHE cosmic radiation.
\bigskip\bigskip

Our research was supported in part by the National Science Foundation
under grant number NSF-PHY/98-02709.

\listrefs
\bye